\documentstyle[12pt]{article}

\begin{document}
\begin{center}
{\bf\large QUANTUM GENERATION OF THE NON-ABELIAN\\
$SU(N)$ GAUGE FIELDS}
\end{center}
\vspace{2mm}
\begin{center}
{\bf P. I. Fomin, T. Yu. Kuzmenko}
\end{center}
\begin{center}
{\sl Bogolyubov Institute for Theoretical Physics,\\
National Academy of Sciences of Ukraine,\\ 
14 b Metrologichna Str., Kyiv-143, 252143, Ukraine\\
e-mail: tanya@ap3.bitp.kiev.ua}
\end{center}

\begin{abstract}
A generation mechanism of the non-Abelian gauge fields in the 
$SU(N)$ gauge theory is investigated. We show that the $SU(N)$ 
gauge fields ensuring the local invariance of the theory are
generated at the quantum level only due to nonsmoothness of 
the scalar phases of the fundamental spinor fields. 
The expressions for the gauge fields are obtained 
in terms of the nonsmooth scalar phases.  
\end{abstract}
\begin{flushleft}
\end{flushleft}

  Nowadays, the gauge principle occupies a significant place in 
quantum field theory. According to this principle, the fundamental 
interactions of elementary particles are transferred by gauge 
fields. The existence of these fields is considered to be 
necessary for ensuring the local gauge symmetries. The local 
$U(1)$ gauge symmetry in quantum electrodynamics was first 
discovered by Weyl [1]. The non-Abelian local gauge symmetries 
and corresponding gauge fields were introduced by Yang and 
Mills [2]. Based on this approach, later on the structure of weak 
and strong interactions was established [3,4].

  It is commonly supposed that the gauge principle must necessarily 
be a consequence of the requirement of the gauge symmetry locality. 
However, it was shown [5] that in the framework of  
classical field theory, the local gauge invariance can be ensured 
without introduction of nontrivial gauge fields, i.e., vector 
fields with nonzero field strengths. It is sufficient to introduce 
only gradient vector field $\partial_{\mu}B(x)$, as a "compensative 
field", with zero strength 
$(\partial_{\mu}\partial_{\nu}-\partial_{\nu}\partial_{\mu})B(x)=0$.
Such field does not contribute to dynamics [5]. From the viewpoint 
of the classification of fields by spin, the scalar field $B(x)$ 
corresponds to spin of zero and gradient vector field
$\partial_{\mu}B(x)$ is longitudinal. True vector gauge fields 
$A_{\mu}$ are transversal fields corresponding to spin of unity. 
Gauge invariance of theory means that the longitudinal part of 
vector gauge fields does not contribute to dynamics.

   If so, what is the real cause of the existence of gauge fields 
and interactions? In Ref.[6], the "quantum gauge principle" was 
formulated in the context of quantum electrodynamics. This principle 
holds that the Abelian $U(1)$ gauge fields are generated at the 
quantum level only and the generation of these fields
is related to nonsmoothness of the field 
trajectories in the Feynman path integrals, by which the field 
quantization is determined. In this paper, we investigate the 
mechanism of non-Abelian $SU(N)$ gauge field generation. It is shown
that the non-Abelian nontrivial vector fields are generated because 
of nonsmoothness of the field trajectories for the scalar phases of 
the spinor fields in the $SU(N)$ gauge theory.

  Let us consider the Lagrangian for free spinor fields
\begin{equation}
L=i\overline{\psi}^{j}{\gamma}^{\mu}\partial_{\mu}{\psi}^{j}
-m\overline{\psi}^{j}{\psi}^{j},
\end{equation}
where $j=1,2,...,N$. In what follows the index $j$ will be omitted.

  The Lagrangian (1) is invariant under global non-Abelian 
$SU(N)$-transformations
\begin{equation} 
{\psi}'(x)=e^{i{t}^a{\omega}_a}{\psi}(x),\ \ \  
\overline{\psi}'(x)=\overline{\psi}(x)e^{-i{t}^a{\omega}_a}, 
\end{equation} 
where $t^a$ are $SU(N)$ group generators, 
$\omega_a=const$, $a=1,2,...,N^2-1$.

  In the framework of classical theory, physical fields are known 
to be described by sufficiently smooth functions. Considering  
smooth local $SU(N)$-transformations
\begin{equation}
{\psi}'(x)=e^{i{t}^a{\omega}_a(x)}{\psi}(x),\ \ \ 
\overline{\psi}'(x)=\overline{\psi}(x)e^{-i{t}^a{\omega}_a(x)},
\end{equation}
we obtain that the transformed Lagrangian differs from the original 
one by the term:
\begin{equation}
\bigtriangleup L=i\overline{\psi}(x)e^{-i{t}^a{\omega}_a(x)} 
{\gamma}^{\mu}(\partial_{\mu}e^{i{t}^b{\omega}_b(x)}){\psi}(x).
\end{equation}
In Ref.[5], it was shown that the local gauge invariance of the
transformed Lagrangian can be ensured by introducing scalar fields
$B_{a}(x)$. To put it another way, the Lagrangian
$$L=i\overline{\psi}{\gamma}^{\mu}\partial_{\mu}{\psi}+
i\overline{\psi}(x)e^{-i{t}^a{B}_a(x)} 
{\gamma}^{\mu}(\partial_{\mu}e^{i{t}^b{B}_b(x)}){\psi}(x)
-m\overline{\psi}{\psi}$$
is invariant under the transformations (3) provided that the fields
$B_{a}(x)$ transform as:
$$e^{i{t}^a{B}'_a(x)}=e^{i{t}^a{B}_a(x)}e^{-i{t}^b{\omega}_b(x)}.$$
The introduced scalar fields $B_{a}(x)$ do not contribute to 
dynamics, since they do not give rise to nonzero strengths and can
be excluded by means of the smooth point transformations of the 
field variables ${\psi}\to \exp\left (i{t}^a{B}_a\right ){\psi}$ 
[5]. Thus we need not compensate the term (4) by introducing 
nontrivial vector fields $A^{a}_{\mu}$ that do not reduce to 
gradients of scalar functions. 

   The situation changes in the quantum approach. In the Feynman 
formulation of quantum field theory the transition amplitudes 
are expressed by the path integrals that are determined
on nonsmooth field trajectories [7]. In this context 
the Lagrangian (1) and its symmetries are determined on the 
class of nonsmooth functions ${\psi}(x)$, corresponding to nonsmooth 
trajectories in path integrals. In the strict sense, the derivatives 
involved in the Lagrangian (1) are discontinuous functions. 
From physics standpoint, field trajectory nonsmoothnesses are 
related to quantum fluctuations of the local fields. Feynman 
integrals, as a rule, are additionally specified by the implicit 
switch to "smoothed-out" approximations [8]. In this case the 
degrees of freedom corresponding to gauge vector fields are lost.
Here we show that, as in quantum electrodynamics [6], in the 
non-Abelian $SU(N)$ gauge theory these degrees of freedom can be 
explicitly taken into account when "smoothing" of nonsmooth fields 
is more carefully carried out.

   Let us approximate nonsmooth functions ${\theta}^{a}(x)$ 
by smooth functions ${\omega}^{a}(x)$:
$$ {\theta}^{a}(x)={\omega}^{a}(x)+...$$
In order to write down the next term of the "smoothed-out" 
representation of the nonsmooth functions ${\theta}^{a}(x)$ it is 
necessary to consider the behaviour of the first derivatives of
${\theta}^{a}(x)$. The derivatives $\partial_{\mu}{\theta}^{a}(x)$ 
at nonsmoothness points of ${\theta}^{a}(x)$ are discontinuous 
functions. Since the derivatives 
$\partial_{\mu}{\omega}^{a}(x)$ are continuous functions,
they approximate badly the behaviour of the derivatives of the 
"smoothed-out" ${\theta}^{a}(x)$. Let us denote a 
difference between them by ${\theta}_{\mu}^{a}(x)$ 
and write $\partial_{\mu}{\theta}^{a}(x)$ as follows:
\begin{equation}
\partial_{\mu}{\theta}^{a}(x)=
\partial_{\mu}{\omega}^{a}(x)+
{\theta}_{\mu}^{a}(x).
\end{equation}
Since the nonsmooth fields ${\theta}_{\mu}^{a}(x)$ do not reduce
to gradients of smooth scalar fields, they are the nontrivial 
vector fields that give rise to nonzero field strengths:
$$\partial_{\mu}{\theta}_{\nu}^{a}(x)-\partial_{\nu}
{\theta}_{\mu}^{a}(x)\neq 0.$$
Therefore the fields $\partial_{\mu}{\theta}^{a}(x)$ 
involve the additional degrees of freedom which are related
to nonsmoothness of the ${\theta}^{a}(x)$. 
It should be noted that the fields ${\theta}_{\mu}^{a}(x)$ are 
ambiguously determined due to ambiguity of choice of 
${\omega}^{a}(x)$.

  On integrating the left and right sides of Eq.(5) over 
space-like contour $(P)$ we obtain:
$$ {\theta}^{a}(x)={\omega}^{a}(x)+\int\limits^{x}_{(P)} dy^{\mu}
{\theta}_{\mu}^{a}(y).$$

  Let us now consider ${\theta}^{a}(x)$ 
as scalar phases of the spinor fields ${\psi}(x)$ realizing the
fundamental representation of the $SU(N)$ gauge group 
and separate out these phase degrees of freedom in 
an explicit form:
\begin{equation} 
{\psi}(x)=e^{i{t}^{a}{\theta}_{a}(x)} {\psi}_{0}(x),
\end{equation}
where the spinor fields ${\psi}_{0}$ are the 
representatives of the class of gauge-equivalent fields [9], 
$e^{i{t}^{a}{\theta}_{a}}$ is a unitary $N\times N$ matrix. 
Then, provided the Lagrangian (1) is determined on the class 
of nonsmooth functions ${\psi}(x)$, using Eq.(6) we obtain:
\begin{equation} 
L=i\overline{\psi}_{0}{\gamma}^{\mu}
\partial_{\mu}{\psi}_{0}+
i\overline{\psi}_{0}
e^{-i{t}^a{\theta}_a}{\gamma}^{\mu}
\left ( \partial_{\mu}e^{i{t}^b{\theta}_b} \right )
{\psi}_{0}-m\overline{\psi}_{0}{\psi}_{0}.
\end{equation}
Represent the matrix $e^{i{t}^{a}{\theta}_{a}}$ as a superposition 
of the unit matrix $I$ and $SU(N)$ group generators $t^{a}$: 
\begin{equation}
e^{i{t}^a{\theta}_a}=CI+iS_{a}t^{a}.
\end{equation}
Since ${t}^{a}$ are traceless matrices normalized by
Tr$({t}^{a}{t}^{b})=\frac{1}{2}{\delta}^{ab}$, the coefficients $C$ 
and $S_{a}$ in Eq.(8) are given by:
\begin{equation}
C=\frac{1}{N}{\mbox{Tr}}\left (
e^{i{t}^a{\theta}_a}\right ),\ \ \
S_{a}=-2i{\mbox{Tr}}\left (
t^{a}e^{i{t}^b{\theta}_b}\right ).
\end{equation}
Then taking into account the commutation rules for $SU(N)$ group 
generators [10] we can write down:
\begin{equation}
e^{-i{t}^{a}{\theta}_{a}}\partial_{\mu}
e^{i{t}^{b}{\theta}_{b}}=
i{t}^{a}\left \{ \bar C\partial_{\mu}S_a-
\bar S_a\partial_{\mu}C+
({f}_{abc}-i{d}_{abc})\bar S^{b}\partial_{\mu}{S}^{c}\right \},
\end{equation}
where ${d}_{abc}$ $({f}_{abc})$ are totally symmetric 
(antisymmetric) structural constants of $SU(N)$-group, the 
overline denotes complex conjugation. It should be noted that 
the terms proportional to the unit matrix are absent in the 
right side of Eq.(10) because 
Tr$ \left (e^{-i{t}^{a}{\theta}_{a}}\partial_{\mu}
e^{i{t}^{b}{\theta}_{b}}\right )=0$.

Since the matrix $e^{i{t}^{a}{\theta}_{a}}$ is unitary, 
the following equation is valid: 
\begin{equation}
\bar C S_{a}-\bar S_{a}C+
({f}_{abc}-i{d}_{abc})\bar S^{b}{S}^{c}=0.
\end{equation}
Differentiating the left and right sides of Eq.(11) and using the
property of antisymmetry of ${f}_{abc}$ we conclude 
that the expression in curly brackets in Eq.(10) is a real function. 
Thus this expression can be identified with the gauge fields: 
\begin{equation}
A_{\mu}^{a}\equiv \bar 
C\partial_{\mu}S^{a}- \bar S^{a}\partial_{\mu}C+ 
({f}^{abc}-i{d}^{abc})\bar S_{b}\partial_{\mu}{S}_{c}.
\end{equation}
Unlike the gauge field in electrodynamics [6], the fields 
$A_{\mu}^{a}$ are nonlinear functions of ${\theta}^{a}(x)$. As a  
consequence of nonsmoothness of the phases ${\theta}^{a}(x)$ the 
fields $A_{\mu}^{a}$ are also not smooth. If we take into account  
only the first term in the right side of relation (5) we obtain that 
the fields $A_{\mu}^{a}$ do not contribute to the dynamics, as in 
classical field theory [5], and the degrees of freedom corresponding 
to gauge vector fields are lost.
The account of ${\theta}_{\mu}^{a}(x)$ enables us to interpret 
the fields $A_{\mu}^{a}$ as 
nontrivial vector fields that give rise to nonzero
field strenghths:
$$\partial_{\mu}A_{\nu}^{a}(x)-\partial_{\nu}A_{\mu}^{a}(x)\neq 0.$$

   By way of illustration let us consider the Yang-Mills $SU(2)$
gauge group. In consequence of anti-commutativity of the $SU(2)$ 
group generators the coefficients $C$ and $S_{a}$ (see Eq.(9))
are given by:
\begin{equation}
C=\cos \left ( {\theta}/{2} \right ), \ \ \
S_{a}=2n_{a}\sin \left ( {\theta}/{2} \right ), 
\end{equation}
where
\begin{equation}
{\theta}=\sqrt{{\theta}_{a}{\theta}^{a}}, \ \ \
n_{a}={\theta}_{a}/{\theta}, \ \ \ a=1,2,3.
\end{equation}
From Eqs.(13) and (14) it follows that the gauge fields 
$A^{a}_{\mu}$ can be written as:
\begin{equation} 
A_{\mu}^{a}= n^a\partial_{\mu}{\theta}+
\sin{\theta}(\partial_{\mu}n^{a})+ 
\sin^{2} ({\theta}/{2}) 
[{\bf n}\times \partial_{\mu}{\bf n}]^{a}.
\end{equation}
Expression (15) demonstrates explicitly the relation between
the Yang-Mills gauge fields and the nonsmooth scalar phases of
the spinor fields.

   Let us obtain the transformation law for the vector fields (12).
For this purpose we consider the infinitesimal smooth local 
transformations for the spinor fields: 
\begin{equation} 
{\psi}'_{0}(x)=e^{i{t}^a{\omega}_a(x)}{\psi}_{0}(x),\ \ \ 
\overline{\psi}'_{0}(x)=\overline{\psi}_{0}(x)
e^{-i{t}^{a}{\omega}_{a}(x)}.
\end{equation}
Then the Lagrangian (7) can be written as:
\begin{equation} 
L=i\overline{\psi}'_{0}{\gamma}^{\mu}
{\partial}_{\mu}{\psi}'_{0}+
i\overline{\psi}'_{0}
e^{i{t}^{a}{\omega}_{a}}
e^{-i{t}^{b}{\theta}_{b}}{\gamma}^{\mu}
\partial_{\mu}\left (e^{i{t}^{c}{\theta}_{c}}
e^{-i{t}^{l}{\omega}_{l}} \right ){\psi}'_{0}-
m\overline{\psi}'_{0}{\psi}'_{0}.
\end{equation}
Defining the gauge fields ${A_{\mu}^{a}}'(x)$ 
similarly to Eqs.(10) and (12) by the following equation:
\begin{equation}
i{t}_{a}{A_{\mu}^{a}}'(x)=
e^{i{t}^{a}{\omega}_{a}}
e^{-i{t}^{b}{\theta}_{b}}\partial_{\mu}\left (
e^{i{t}^{c}{\theta}_{c}}
e^{-i{t}^{l}{\omega}_{l}}\right ),
\end{equation}
we find that the transformed gauge fields
${A_{\mu}^{a}}'(x)$ are related to the fields (12) as follows:
\begin{equation}
{A_{\mu}^{a}}'(x)=A_{\mu}^{a}(x)-{\partial}_{\mu}{\omega}^{a}(x)-
f_{abc}{\omega}^{b}(x)A_{\mu}^{c}(x).
\end{equation}
Consequently, in the framework of considered scheme of the gauge
field generation we derive the usual transformation law for the
$SU(N)$ gauge fields, with the local gauge invariance of the 
Lagrangian (7) being not necessary. 

   Using Eqs.(10) and (12) we obtain that the Lagrangian (7)
takes the form:
\begin{equation} 
L=i\overline{\psi}_{0}{\gamma}^{\mu}\hat D_{\mu}{\psi}_{0}-
m\overline{\psi}_{0}{\psi}_{0},
\end{equation}
where $\hat D_{\mu}\equiv \partial_{\mu}+iA_{\mu}^{a}{t}_{a}$ is 
the covariant derivative. It is easy to verify that the 
Lagrangian (20) is invariant under the transformations 
(16) and (19).

  Therefore the gauge fields $A_{\mu}^{a}$ ensuring the local
$SU(N)$ gauge invariance of the Lagrangian (20) are generated 
because of nonsmoothness of the field trajectories in Feynman path 
integral. The nonsmoothness of the fields $A_{\mu}^{a}$ 
corresponds to their quantum nature and means that these fields 
should also be quantized, i.e., continual integration is to be 
carried out over the variables $A_{\mu}^{a}(x)$.  However the 
fields $A_{\mu}^{a}$ in the Lagrangian (20) do not exhibit all the 
properties of physical fields since they cannot propagate in space 
because of the absence of the kinetic term. 

  An  expression similar to the kinetic term can be obtained by the 
calculation of the effective action for the spinor fields described 
by the Lagrangian (20). Using the results of the calculations 
performed in Ref.[11], we find the following expression for the 
kinetic term in the one-loop approximation 
\begin{equation}
L_{\mbox{\small{eff}}}=\kappa 
\ln \frac{\Lambda}{{\mu}_0}tr\hat F_{\mu\nu}^2,\ \ \
\hat F_{\mu\nu}=[\hat D_{\mu},\hat D_{\nu}],
\end{equation}
where ${\Lambda}$ and  ${\mu}_{0}$ are the momentum of the 
ultraviolet and infrared cut-off respectively; ${\kappa}$ is the 
numerical coefficient. 

  The formula (21) takes the usual form [10]
\begin{displaymath} 
L_{\mbox{\small{eff}}}=\frac {\hbar c}{8g^2}
{\mbox{tr}}F_{\mu\nu}^2
\end{displaymath} 
upon identifying
\begin{equation}
g^2=\frac {\hbar c}{8\kappa \ln \frac {\Lambda}{{\mu}_0}}.
\end{equation} 
The last equation relates the charge $g$ with the parameters 
${\Lambda}$ and ${\mu}_0$ as well as with the world's constants  
${\hbar}$ and $c$, and thus demonstrates explicitly quantum 
origin of the charge. 

  We note in conclusion that the "compensating" gauge fields need 
not be artificially introduced for the local gauge invariance of 
the theory to be ensured. The vector gauge fields are generated 
through nonsmoothness of the scalar phases of the fundamental 
spinor fields. From the viewpoint of the described scheme of the 
gauge field generation, the gauge principle is an "automatic" 
consequence of field trajectory nonsmoothness in Feynman path 
integral.  
\begin{center}
{\bf Acknowledgement}
\end{center}

  This work is supported in part by Swiss National Science 
Foundation Grant CEEC/NIS/96-98/7 IP 051219. One of the authors 
(PIF) is thankful to Professor H. Leutwyler for the kind hospitality
at ITP of Bern University. We would like to thank Yu. Shtanov for 
several helpful comments and a careful reading of the manuscript.

\end{document}